\documentstyle[prb,aps,twocolumn,epsfig,floats,amssymb,amsmath]{revtex}

\newcommand  {\rsq}    {\left< \vec r^{2} \right>}

\newcommand  {\Ham}    {{\cal H}}
\newcommand  {\ij}    {{\left<i,j\right>}}
\newcommand  {\BoxT}    {{\tensor\lambda}}

\newcommand  {\Mvec}[1]    {\underline{#1}\,}
\newcommand  {\Mtensor}[1]    {\underline{\underline{#1}}\,}
\newcommand  {\MTvec}[1]    {\underline{\tilde{#1}}\,}
\newcommand  {\MTtensor}[1]    {\underline{\underline{\tilde{#1}}}\,}

\begin{document}
\draft\title{Tube Models for
Rubber-Elastic Systems}

\author{Boris Mergell and Ralf Everaers*}

\address{Max-Planck-Institut f\"ur Polymerforschung,
  Postfach 3148, D-55021 Mainz, Germany}

\address{
\begin{minipage}{5.55in}
\begin{abstract}\hskip 0.15in
  In the first part of the paper we show that the constraining
  potentials introduced to mimic entanglement effects in Edward's tube
  model and Flory's constrained junction model are diagonal in the
  generalized Rouse modes of the corresponding phantom network. As a
  consequence, both models can formally be solved exactly for
  arbitrary connectivity using the recently introduced constrained
  mode model.  In the second part, we solve a double tube model for
  the confinement of long paths in polymer networks which is partially
  due to crosslinking and partially due to entanglements. Our model
  describes a non-trivial crossover between the Warner-Edwards and the
  Heinrich-Straube tube models. We present results for the macroscopic
  elastic properties as well as for the microscopic deformations
  including structure factors.
\end{abstract}
\end{minipage}
\vspace*{-0.5cm} 
} 

\maketitle

\section{Introduction}

Polymer networks~\cite{Treloar75} are the basic structural element of
systems as different as tire rubber and gels and have a wide range of
technical and biological applications. From a macroscopic point of
view, rubber-like materials have very distinct visco- and
thermoelastic properties.~\cite{Treloar75,DoiEdwards_86} They
reversibly sustain elongations of up to 1000\% with small strain
elastic moduli which are four or five orders of magnitude smaller than
for other solids.  Maybe even more unusual are the thermoelastic
properties discovered by Gough and Joule in the 19th century: when
heated, a piece of rubber under a constant load contracts, and,
conversely, heat is released during stretching. This implies that the
stress induced by a deformation is mostly due to a decrease in
entropy.  The microscopic, statistical mechanical origin of this
entropy change remained obscure until the discovery of polymeric
molecules and their high degree of conformational flexibility in the
1930s.  In a melt of identical chains polymers adopt random coil
conformations~\cite{Flory_jcp_49} with mean-square end-to-end
distances proportional to their length, $\rsq\sim N$. A simple
statistical mechanical argument, which only takes the connectivity of
the chains into account, then suggests that flexible polymers react to
forces on their ends as linear, {\em entropic springs}. The spring
constant, $k = \frac{3 k_B T}{\rsq}$, is proportional to the
temperature. Treating a piece of rubber as a random network of
non-interacting entropic springs (the phantom
model~\cite{James_jcp_47,JamesGu_jcp_47,Flory_procroy_76})
qualitatively explains the observed behavior, including --- to a first
approximation --- the shape of the measured stress-strain curves.

In spite of more than sixty years of growing qualitative
understanding, a rigorous statistical mechanical treatment of polymer
networks remains a challenge to the present day. Similar to spin
glasses,~\cite{MezardParisi_87} the main difficulty is the presence of
quenched disorder over which thermodynamic variables need to be
averaged. In the case of polymer
networks,~\cite{Edwards_procphyssoc_67,Edwards_jphysA_68,DeamEdw_philtrans_76}
the vulcanization process leads to a simultaneous quench of two
different kinds of disorder: (i) a random connectivity due to the
introduction of chemical crosslinks and (ii) a random topology due to
the formation of closed loops and the mutual impenetrability of the
polymer backbones. Since for instantaneous crosslinking
monomer-monomer contacts and entanglements become quenched with a
probability proportional to their occurrence in the melt, {\em
  ensemble} averages of {\em static} expectation values for the chain
structure etc. are {\em not} affected by the vulcanization as long as
the system remains in its state of preparation.

For a given connectivity the phantom model Hamiltonian for
non-interacting polymer chains formally takes a simple quadratic
form,~\cite{James_jcp_47,JamesGu_jcp_47,Flory_procroy_76} so that one
can at least formulate theories which take the random connectivity of
the networks fully into
account.~\cite{Eichinger_mm_72,HiggsBall_jpf_88,Zippelius_epl_93} The
situation is less clear for entanglements or topological constraints,
since they do not enter the Hamiltonian as such, but divide phase
space into accessible and inaccessible regions. In simple cases,
entanglements can be characterized by topological invariants from
mathematical knot
theory.~\cite{Edwards_procphyssoc_67,Edwards_jphysA_68} However,
attempts to formulate topological theories of rubber elasticity (for
references see~\cite{RE_njp_99}) encounter serious difficulties.  Most
theories therefore omit such a detailed description in favor of a
mean-field ansatz where the different parts of the network are thought
to move in a deformation-dependent elastic matrix which exerts
restoring forces towards some rest positions.  These restoring forces
may be due to chemical crosslinks which localize random paths through
the network in space~\cite{WarnerEdwards_jpa_78} or to entanglements.
The classical theories of rubber elasticity~\cite{Treloar75,RoncaAll_jcp_75,Flory_jcp_77,%
  ErmanFlory_jcp_78,FloryErman_mm_82,Kaest_collpolsc_81} assume that
entanglements act only on the crosslinks or junction points, while the
tube models~\cite{DoiEdwards_86,Edwards_tube_procphyssoc_67,Marrucci_mm_81,Graessley_advpol_82,Gaylord_polbull_82,%
  Heinrich_advpol_88,EdwardsVil_reppro_88} stress the importance of
the topological constraints acting along the contour of strands
exceeding a minimum ``entanglement length'', $N_e$. Originally devised
for polymer networks, the tube concept is particularly successful in
explaining the extremely long relaxation times in non-crosslinked
polymer melts as the result of a one-dimensional, curvilinear
diffusion called reptation~\cite{deGennes_jcp_71} of linear chains of
length $N\gg N_e$ within and finally out of their original tubes.
Over the last decade computer
simulations~\cite{RE_njp_99,Dueringkkgsg_Prl_91,DueringKKGSG_jcp_94,REKK_mm_95}
and experiments~\cite{Straube_mm_94,Straube_prl_95,Straube_mm_96} have
finally also collected mounting evidence for the importance and
correctness of the tube concept in the description of polymer
networks.

More than thirty years after its introduction and in spite of its
intuitivity and its success in providing a unified view on entangled
polymer networks and
melts,~\cite{DoiEdwards_86,Graessley_advpol_82,Heinrich_advpol_88,EdwardsVil_reppro_88}
there exists to date no complete solution of the Edwards tube model
for polymer networks. Some of the open problems are apparent from a
recent controversy on the interpretation of SANS
data.~\cite{Straube_prl_95,Straube_mm_96,ReadMcLeish_prl_97,ReadMcLeish_mm_97,Straube_prl_98,Read_prl_98}
Such data constitute an important experimental test of the tube
concept, since they contain information on the degree and deformation
dependence of the confinement of the microscopic chain motion and
therefore allow for a more detailed test of theories of rubber
elasticity than rheological
data.~\cite{Gottlieb_pol_83,Gottlieb_mm_84,Gottlieb_mm_87}

On the theoretical side, the original approach of Warner and
Edwards~\cite{WarnerEdwards_jpa_78} used mathematically rather
involved replica methods~\cite{EdwardsVil_reppro_88} to describe the
localization of a long polymer chain in space due to crosslinking. The
replica method allows for a very elegant, self-consistent introduction
of constraining potentials, which confine {\em individual} polymer
strands to random-walk like tubular regions in space while {\em
  ensemble} averages over all polymers remain identical to those of
unconstrained chains. Later Heinrich and
Straube~\cite{Heinrich_advpol_88,Straube_prl_95} recalculated these
results for a solely entangled system where they argued that there are
qualitative differences between confinement due to entanglements and
confinement due to crosslinking. In particular, they argued that the
strength of the confining potential should vary affinely with the
macroscopic strain, resulting in fluctuations perpendicular to the
tube axis which vary only like the square root of the macroscopic
strain.

Replica calculations provide limited insight into physical mechanism
and make approximations which are difficult to
control.~\cite{HiggsBall_jpf_88} It is therefore interesting to note
that Flory was able to solve the, in many respects similar,
constrained-junction model~\cite{Flory_jcp_77} without using such
methods. Recent refinements of the constrained-junction model such as
the constrained-chain model~\cite{ErmanMon_mm_89} and the
diffused-constrained model~\cite{KlocMark_mm_95} have more or less
converged to the (Heinrich and Straube) tube model, even though the
term is not mentioned explicitly. Another variant of this model was
recently solved by Rubinstein and Panyukov.~\cite{RubinPan_mm_97} In
particular, the authors illustrated how non-trivial, sub-affine
deformations of the polymer strands result from an affinely deforming
confining potential.

While tube models are usually formulated and discussed in real space,
two other recent papers have pointed independently to considerable
simplifications of the calculations in mode space. Read and
McLeish~\cite{ReadMcLeish_mm_97} were able to rederive the
Warner-Edwards result in a particularly simple and transparent manner
by showing that a harmonic tube potential is diagonal in the Rouse
modes of a linear chain. Complementary, one of the present authors
introduced a general constrained mode model (CMM),~\cite{RE_epjb_98}
where confinement is modeled by deformation dependent linear forces
coupled to (approximate) {\em eigenmodes} of the phantom network
instead of a tube-like potential in real space.  This model can easily
be solved exactly and is particularly suited for the analysis of
simulation data, where its parameters, the degrees of confinement for
all considered modes, are directly measurable.  Simulations of
defect-free model polymer networks under strain analyzed in the
framework of the CMM~\cite{RE_njp_99} provide evidence that it is
indeed possible to predict {\em macroscopic} restoring forces and {\em
  microscopic} deformations from constrained fluctuation theories.  In
particular, the results support the choice of
Flory,~\cite{Flory_jcp_77} Heinrich and
Straube,~\cite{Heinrich_advpol_88} and Rubinstein and
Panyukov~\cite{RubinPan_mm_97} for the deformation dependence of the
confining potential. In spite of this success, the CMM in its original
form suffers from two important deficits: (i) due to the multitude of
independent parameters it is completely useless for a comparison to
experiment and (ii) apart from recovering the tube model on a scaling
level, Ref.~\cite{RE_epjb_98} remained fairly vague on the exact
relation between the approximations made by the Edwards tube model and
the CMM respectively.

In the present paper we show that the two models are, in fact,
equivalent. The proof, presented in section~\ref{sec:ch} is a
generalization of the result by Read and McLeish to arbitrary
connectivity.  It provides the link between the considerations of
Eichinger,\cite{Eichinger_mm_72} Graessley,~\cite{Graessley_mm_75}
Mark,~\cite{KlocMark_mm_90} and others on the dynamics of (micro)
phantom networks and the ideas of Edwards and Flory on the suppression
of fluctuations due to entanglements.  As a consequence, the CMM can
be used to formally solve the Edwards tube model exactly, while in
turn the independent parameters of the CMM are obtained as a function
of a single parameter: the strength of the tube potential.  Quite
interestingly, it turns out that the entanglement contribution to the
shear modulus depends on the connectivity of the network. In order to
explore the consequences, we discuss in the second part the
introduction of entanglement effects into the Warner-Edwards model,
which represents the network as an ensemble of independent long paths
comprising many strands. Besides recovering some results by Rubinstein
and Panyukov for entanglement dominated systems, we also calculate the
single chain structure factor for this controversial
case.~\cite{Straube_prl_95,Straube_mm_96,ReadMcLeish_prl_97,ReadMcLeish_mm_97,Straube_prl_98,Read_prl_98}
Finally we propose a ``double tube'' model to describe systems where
the confinement of the fluctuations due to crosslinks and due to
entanglements is of similar importance and where both effects are
treated within the same formalism.

\section{Constrained fluctuations in networks of arbitrary connectivity}
\subsection{The Phantom Model}
\label{sec:PhantomModel}
The Hamiltonian of the phantom
model~\cite{James_jcp_47,JamesGu_jcp_47,Flory_procroy_76} is given by
$\Ham_{ph} = \frac k2 \sum_\ij^{M} r_{ij}^2$, where $\ij$ denotes a
pair of nodes $i,j \in {1,\ldots,M}$ which are connected by a polymer
chain acting as an entropic spring of strength $k = \frac{3 k_B
  T}{\rsq}$, and $\vec r_{ij}(t) = \vec r_i(t) - \vec r_j(t)$ the
distance between them. In order to simplify the notation, we always
assume that all elementary springs have the same strength $k$.  The
problem is most conveniently studied using periodic boundary
conditions, which span the network over a fixed
volume~\cite{DeamEdw_philtrans_76} and define the equilibrium position
$\vec R_i = (X_i,Y_i,Z_i)$. A conformation of a network of harmonic
springs can be analyzed in terms of either the bead positions $\vec
r_i(t)$ or the deviations $\vec u_i(t)$ of the nodes from their
equilibrium positions $\vec R_i$.  In this representation, the
Hamiltonian separates into two independent contributions from the
equilibrium extensions of the springs and the fluctuations.  For the
following considerations it is useful to write fluctuations as a
quadratic form.~\cite{Eichinger_mm_72} Finally, we note that the
problem separates in Cartesian coordinates $\alpha=x,y,z$ due to the
linearity of the springs.  In the following we simplify the notation
by writing the equations only for one spatial dimension:

\begin{equation}
\Ham_{ph} =  \frac k2 \sum_\ij X_{ij\alpha}^2 + 
 \frac 12\, \Mvec{u}^{t} \Mtensor{K} \Mvec{u}.
\label{eq:Hph_pos}
\end{equation}
Here ${\Mvec u}$ denotes a $M$-dimensional vector 
with $(\Mvec u)_i \equiv (\vec u_i)_x$.
$\Mtensor{K}$ is the connectivity or Kirchhoff matrix whose diagonal
elements $(\Mtensor{K})_{ii} = f_i k$ are given by the node's functionality
(e.g. a node which is part of a linear chain is connected
to its two neighbors so that $f_i=2$ in contrast to a 
four-functional crosslink with $f_i=4$). The off-diagonal
elements of the Kirchhoff matrix are given by $(\Mtensor{K})_{ij} = -k$,
if nodes $i$ and $j$ are connected and by $(\Mtensor{K})_{ij} = 0$
otherwise. Furthermore, we have assumed that all network strands
have the same length.

The fluctuations can be written as a sum
over independent modes $\Mvec{e}_p$ which are the eigenvectors
of the Kirchhoff matrix: $\Mtensor{K} \Mvec{e}_p = k_p \Mvec{e}_p$
where the $\Mvec{e}_p$ can be chosen to be orthonormal
$\Mvec{e}_p \cdot \Mvec{e}_{p\prime} = \delta_{pp\prime}$.
The transformation to the eigenvector representation 
$\MTvec{u} = \Mtensor{S}\Mvec{u}$ and back to the node representation
$\Mvec{u} = \Mtensor{S}^{-1}\MTvec{u}$ is mediated by a matrix
$\Mtensor{S}$ whose column vectors correspond to the $\Mvec{e}_p$.
By construction, $\Mtensor{S}$ is orthogonal with 
$\Mtensor{S}^t=\Mtensor{S}^{-1}$.
Furthermore, the Kirchhoff matrix is diagonal in the 
eigenvector representation
$(\MTtensor{K})_{pp} = (\Mtensor{S}^{-1}\Mtensor{K}\Mtensor{S})_{pp} = k_p$.
The Hamiltonian then reduces to
\begin{equation}
\Ham_{ph} =  \frac k2 \sum_\ij X_{ij}^2 + 
\sum_p  \frac {k_p}2 {\tilde u_{p}}^2.
\label{eq:Hph_modes}
\end{equation}
Since the connectivity is the result of a random process, it is
difficult to discuss the properties of the Kirchhoff matrix and the
eigenmode spectrum in general.~\cite{Eichinger_mm_72,Graessley_mm_75}
The following simple argument~\cite{RE_epjb_98} ignores these
difficulties. The idea is to relate the mean square equilibrium
distances $\langle X_{ij}^2 \rangle$ to the thermal fluctuations of
the phantom network.

Consider the network strands before and after the formation of
the network by end-linking. In the melt state, the typical
mean {\em square} extension $\rsq$ is entirely due to thermal fluctuations,
while $\langle\vec r\rangle = 0$. In the crosslinked state, the strands
show reduced thermal fluctuations $\langle \vec u_{ij}^2\rangle$
around {\em quenched}, non-vanishing mean extensions $\langle \vec R_{ij}^2\rangle$.
However, the ensemble average of the total extension 
$\langle \vec R_{ij}^2\rangle + \langle \vec u_{ij}^2\rangle$
is {\em not} affected by the end-linking procedure. 
The fluctuation contribution $\langle \vec u_{ij}^2\rangle$ depends
on the connectivity of the network and can be estimated using
the equipartition theorem.    
The total thermal energy in the fluctuations, $U_{fluc}$, is
given by $\frac32 k_BT$ times the number of modes
and therefore $U_{fluc} = \frac32 k_BT N_{nodes} = 
\frac2f \frac32 k_BT N_{strands}$, where $N_{nodes}$ 
and $N_{strands}$ are the number of junction points
and network strands, which are related by $N_{strands}=\frac f2 N_{nodes}$
in an $f$-functional network. Equating the thermal energy
per mode to $\frac k2 \langle{\vec u_{ij}}^2\rangle$, one 
obtains~\cite{Flory_procroy_76,RE_epjb_98,Graessley_mm_75}

\begin{eqnarray}
 \langle \vec u_{ij}^2 \rangle &=& \frac2f \rsq\ \ \ \ 
\label{eq:u_ij^2}\\
 \langle \vec R_{ij}^2 \rangle &=& \left(1-\frac2f\right) \rsq\ \ \ \ .
\label{eq:R_ij^2}
\end{eqnarray}

Using these results, one can finally estimate the elastic properties
of randomly cross- or endlinked phantom networks.  Since the
fluctuations are independent of size and shape of the network, they do
not contribute to the elastic response.
The equilibrium positions of the junction points, on the other hand, 
change affinely in the macroscopic strain. The
elastic free energy density due to a volume-conserving, uni-axial
elongation with $\lambda_{||} = \lambda_\perp^{-1/2} = \lambda$
is simply given by:
\begin{eqnarray}
\label{eq:Fph}
\Delta F_{ph}(\lambda) &=&  
 \left(\lambda^2 +\frac 2\lambda - 3\right) 
 \frac{\langle R^2_{strand}\rangle}\rsq\ \rho_{strand}\nonumber\\
&=& 
 \left(\lambda^2 +\frac 2\lambda - 3\right) \ 
 \left(1-\frac2f\right)\ \rho_{strand}\ ,
\end{eqnarray}
where $\rho_{strand}$ is the number density of elastically active strands.
For incompressible materials such as rubber, the shear modulus is
given by $\frac{1}{3}$ of the second derivative of the corresponding
free energy density with respect to the strain parameter $\lambda$. In
response to a finite strain, the system develops a normal tension
$\sigma_{T}$:

\begin{eqnarray}
G_{ph} &=&
\frac{1}{V}\frac{1}{3} \frac{d^2 \Delta F_{ph}(\lambda)}{d\lambda^2}\Bigg\vert_{\lambda=1}\\
&=& \left(1-\frac2f\right)\ \rho_{strand}k_BT\label{eq:Gph}\\
\sigma_{T} &=& (\lambda^2-\frac1\lambda) G_{ph}\label{eq:sigma_T}\ .
\end{eqnarray}
Experimentally observed stress-strain curves show deviations
from eq~\ref{eq:sigma_T}. Usually the results are normalized to
the classical prediction and plotted versus the inverse strain
$1/\lambda$, since they often follow the semi-empirical
Mooney-Rivlin form

\begin{eqnarray}
\frac{\sigma_{T}}{\lambda^2-\frac1\lambda} &\approx& 2C_1 + \frac{2C_2}\lambda
\ \ .\label{eq:mooney}
\end{eqnarray}

\subsection{The Constraint Hamiltonian} \label{sec:ch}

Most theories introduce the entanglement effects as additional,
single-node terms into the phantom model Hamiltonian, which constrain
the movement of the monomers and junction points.  The standard choice
are anisotropic, harmonic springs of strength $\tensor{l}(\lambda)$
between the nodes and points $\vec \xi_i(\lambda)$ which are fixed in
space:
\begin{equation}
\Ham_{constr} = \sum_{i} 
\frac 12 \left(\vec r_i - \vec \xi_i(\lambda)\right)^t \ 
\tensor{l}(\lambda)\ \left(\vec r_i - \vec \xi_i(\lambda)\right)
\label{eq:HamConstHarm}\ \ \ .
\end{equation}
While all models assume that the tube position changes affinely with
the macroscopic deformation,
\begin{eqnarray}  \label{eq:AffTrans}
\vec \xi_i(\lambda)=\BoxT \vec \xi_i(\lambda=1)\ \ ,
\end{eqnarray}
there are two different choices for the deformation dependence
of the confining potential:
\begin{eqnarray}  
\label{eq:l_const}
   \tensor{l}(\lambda)= \tensor{l}(\lambda=1)\\
\label{eq:l_affine}
   \tensor{l}(\lambda)=\BoxT^{-2} \tensor{l}(\lambda=1).
\end{eqnarray}
Since this choice of $\Ham_{constr}$ leaves the different spatial
dimensions uncoupled, we consider the problem again in one dimension and
express $\Ham_{constr}$ in the eigenvector
representation of the Kirchhoff matrix of the unconstrained
network. Using $\Mvec{v}(\lambda) = \Mvec{\xi_x}(\lambda)-\Mvec{X}(\lambda)
= \BoxT \Mvec{v}(\lambda=1)$ one obtains
\begin{equation}
  \begin{split}
    \Ham_{constr} &= \frac {l(\lambda)}2 (\Mvec{u}-\Mvec{v})^{t} (\Mvec{u}-\Mvec{v})\\
    &= \frac {l(\lambda)}2 (\Mtensor{S}^{-1}\MTvec{u}-\Mtensor{S}^{-1}\MTvec{v})^{t} 
    (\Mtensor{S}^{-1}\MTvec{u}-\Mtensor{S}^{-1}\MTvec{v})\\
    &= \frac {l(\lambda)}2 (\MTvec{u}-\MTvec{v})^{t}\Mtensor{S}\Mtensor{S}^{-1} (\MTvec{u}-\MTvec{v})\\
    &= \frac {l(\lambda)}2 (\MTvec{u}-\MTvec{v})^{t} (\MTvec{u}-\MTvec{v})\\
    &= \sum_p  \frac {l(\lambda)}2 \left( {\tilde u_{p}-\tilde v_{p}}\right)^2.
    \label{eq:HamConstMode}
  \end{split}
\end{equation}
Thus the introduction of the single node springs does not change the
eigen{\em vectors} of the original Kirchhoff matrix. The derivation of
eq~\ref{eq:HamConstMode}, which is the Hamiltonian of the Constrained
Mode Model (CMM),~\cite{RE_epjb_98} is a central result of this work.
It provides the link between the considerations of
Eichinger,~\cite{Eichinger_mm_72} Graessley,~\cite{Graessley_mm_75}
Mark,~\cite{KlocMark_mm_90} and others on the dynamics of (micro)
phantom networks and the ideas of Edwards and Flory on the suppression
of fluctuations due to entanglements.

\subsection{Solution and Disorder Averages: The Constrained Mode Model (CMM)}

Since the total Hamiltonian of the CMM
\begin{equation}
  {\cal H}={\cal H}_{ph}+{\cal H}_{constr}
\end{equation}
is diagonal and quadratic in the modes, both the exact solution of the
model for given $\vec{v}_p$ and the {\em subsequent} calculation of
averages over the quenched Gaussian disorder in the $\vec{v}_p$ are
extremely simple.~\cite{RE_epjb_98} In the following we summarize the
results and give general expressions for quantities of physical
interest such as shear moduli, stress-strain relations, and
microscopic deformations.

Consider an arbitrary mode $\vec u_p$ of the polymer network.  Under
the influence of the constraining potential, each Cartesian component
$\alpha$ will fluctuate around a non-vanishing mean excitation
$\vec{U}_p$ with
\begin{equation} \label{U_p}
  U_{p\alpha}(\lambda)=\frac{l_{\alpha}(\lambda)}{k_p+l_{\alpha}(\lambda)}\ v_{p\alpha}(\lambda).
\end{equation}
Using the notation $\vec{\delta u}_p \equiv \vec u_p - \vec U_p$,
 the Hamiltonian for this mode reads
\begin{equation} \label{H_p}
  \begin{split}
    \Ham_{p\alpha}[v_{p\alpha}] =&
    \frac{k_p}{2}U_{p\alpha}^2(\lambda) 
    + \frac{l_{\alpha}(\lambda)}{2}\left(U_{p\alpha}(\lambda)-v_{p\alpha}(\lambda)\right)^2   \\
    &+ \frac{l_{\alpha}(\lambda)+k_p}{2}\ \delta{u}_{p\alpha}^2.
  \end{split}
\end{equation}
Expectation values are calculated by averaging over both
the thermal and the static fluctuations, which are due to the
quenched {\em topological} disorder:\footnote{In order to
simplify the notation, we use $l\equiv l(\lambda=1)$, 
$\langle v_{p\alpha}^2\rangle \equiv 
\langle v_{p\alpha}^2(\lambda=1)\rangle$ etc.}
\begin{equation}  
  \langle A_p(\lambda) \rangle =
  \int dv_{p} \int d \delta u_{p}\  A_{p}[v_{p},\delta u_{p}]\ 
  P(v_{p\alpha})P(\delta u_{p\alpha}) .
\end{equation}
Both distributions are Gaussian and their widths
\begin{eqnarray}  
  \label{deltau_p}
  \langle \delta u^2_{p\alpha}(\lambda)\rangle &=& 
  \frac{k_BT}{k_p+l_{\alpha}(\lambda)}\\
  \label{v_p}
  \langle v_{p\alpha}^2\rangle &=& \frac{k_p+l}{k_p l}k_BT
\end{eqnarray}
follow from the Hamiltonian and the condition that the random
introduction of topological constraints on the {\em dynamics} does
not affect {\em static} expectation values {\em in the state of 
preparation}. In particular,
\begin{equation} \label{statik}
  \langle u_{p\alpha}^2 \rangle = 
  \langle{\delta u}^2_{p\alpha} \rangle + 
  \langle U_{p\alpha}^2\rangle  \equiv \frac{k_BT}{k_p}.
\end{equation}
Eq~\ref{statik} relates the strength $l$ of the confining
potential to the width of $P({v_{p\alpha}})$. The result,
$ \langle v_{p\alpha}^2\rangle = \frac{1}{\gamma_{p}}\frac{k_BT}{k_p}$,
$ \langle U_{p\alpha}^2\rangle = \gamma_{p} \frac{k_BT}{k_p}$,
$ \langle{\delta u}_{p\alpha}^2\rangle = (1-\gamma_{p})\frac{k_BT}{k_p}$
can be expressed conveniently using a parameter 
\begin{equation}
  \label{eq:gamma_p}
  0\le \gamma_{p}\equiv \frac{l}{k_p+l}\le 1,
\end{equation}
which measures the degree of confinement of the modes.
As a result, one obtains for the mean square static excitations:
\begin{equation}
  \label{eq:U_p}
  \langle U_{p\alpha}^2(\lambda)\rangle =
  \lambda_\alpha^2 \left(\frac{l_{\alpha}(\lambda)}
    {k_p+l_{\alpha}(\lambda)} \right)^2 
  \frac{k_p+l}{l} \frac{k_BT}{k_p}\ \ \ .
\end{equation}

Quantities of physical interest are typically sums over the eigenmodes
of the Kirchhoff matrix. For example, the tube diameter is defined as
the average width of the thermal fluctuations of the nodes:
\begin{equation}
  d_{T\alpha}^2(\lambda) = 
  \frac 1M \sum_p \langle\delta u_{p\alpha}^2(\lambda)\rangle \ .
\end{equation}
In particular,
\begin{equation}
  \label{eq:dT}
  d_{T\alpha}^2 = 
  \frac {k_BT}{M\, l} \sum_p \gamma_p.
\end{equation}
More generally, distances between any two monomers 
$r_{nm\alpha}=r_{n\alpha}-r_{m\alpha}$ in real space
are given by
\begin{equation} \label{eq:dist}
  \langle r^2_{nm\alpha}(\lambda) \rangle = 
  \sum_p \langle u^2_{p\alpha}(\lambda) \rangle S^2_{p,nm}
  + \lambda_\alpha^2 R^2_{nm\alpha}\ .
\end{equation}

For the discussion of the elastic properties of the different tube
models it turns out to be useful to define the sum
\begin{equation}\label{eq:g}
  g(\lambda) =
  \frac{k_BT}{V} \frac1{1-\lambda_\alpha^2}\sum_p
  \left(\frac{\langle u_{p\alpha}^2\rangle(\lambda)}{\langle u_{p\alpha}^2\rangle} -1 \right).
\end{equation}
Using eq~\ref{eq:g} the confinement contribution to the normal
tension~\cite{DoiEdwards_86,RE_epjb_98} and the shear modulus can be
written as
\begin{eqnarray}
  \label{eq:sigma}
  \sigma_T(\lambda) &=& \frac1V \sum_p k_p 
  \left( \langle u_{p||}^2(\lambda)\rangle -
    \langle u_{p\perp}^2(\lambda)\rangle \right)\nonumber\\
  &=&
  \left( \lambda^2 - 1\right) g(\lambda) +
  \left( 1-\lambda^{-1}\right) g(\lambda^{-1/2})\\
  \label{eq:G}
  G_{constr} &=& g(1).
\end{eqnarray}

\subsection{Model A: Deformation independent strength of the 
Confining Potential}

In order to completely define the model, one needs to specify the
deformation dependence of the confining potential.
One plausible choice is
\begin{equation}  \label{eq:l_indep}
  \tensor{l}_{\!A}(\lambda)= \tensor{l}_{\!A}(\lambda=1)\ ,
\end{equation}
i.e. a confining potential whose {\em strength}
is strain independent. The following discussion will make clear,
that this choice leads to a situation which mathematically resembles
the phantom model without constraints.

Using eq~\ref{eq:l_indep} the thermal fluctuations 
(and therefore also the tube diameter eq~\ref{eq:dT}) are deformation
independent and remain isotropic in strained systems.
The mean excitations, on the other hand, vary affinely
with the macroscopic strain. This leads to the following
relation for the deformation dependence of the 
total excitation of the modes:
\begin{equation}\label{eq:u2_lambda_A}
  \frac{\langle u_{p\alpha}^2\rangle(\lambda)}{\langle u_{p\alpha}^2\rangle} = 1 +
  (\lambda_{\alpha}^2-1)\frac{l_A}{k_p+l_A}.
\end{equation}
Using eqs~\ref{eq:g} to~\ref{eq:G} one obtains via
\begin{equation}
  g_A(\lambda) = 
  \frac{k_BT}{V}\sum_p
  \left( \frac{l_A}{k_p + l_A}\right).
\end{equation}
a classical stress-strain relation:
\begin{eqnarray}
  \label{eq:sigmaA}
  \sigma_T(\lambda) &=& 
  \left( \lambda^2 - \lambda^{-1}\right) G_A\\
  \label{eq:GA}
  G_{A} &=& \frac{k_BT}{V}\sum_p \gamma_p.
\end{eqnarray}

\subsection{Model B: Affine deformation of the confining potential}

The ansatz 
\begin{eqnarray}  \label{eq:l_AffTrans}
  \tensor{l}_{\!B}(\lambda)=\BoxT^{-2}\ \tensor{l}_{\!B}(\lambda=1)\ \ ,
\end{eqnarray}
goes back to Ronca and Allegra~\cite{RoncaAll_jcp_75} and was used by
Flory, Heinrich and Straube~\cite{Heinrich_advpol_88} and Rubinstein
and Panyukov.~\cite{RubinPan_mm_97} It corresponds to affinely
deforming cavities and leads to a more complex behavior including
corrections to the classically predicted stress-strain behavior.

Using eq~\ref{eq:l_AffTrans}, the mean excitations of partially
frozen modes as well as the thermal fluctuations, become deformation
dependent. The total excitation of a mode is given by:
\begin{equation} \label{eq:u2_lambda_B}
  \frac{\langle u_{p\alpha}^2\rangle(\lambda)}
  {\langle u_{p\alpha}^2\rangle}  = 
  1 + \left( \lambda_\alpha^2 - 1\right)
  \left( \frac{l_B(\lambda)}{k_p + l_B(\lambda)}\right)^2.
\end{equation}
Only in the limit of completely frozen modes, $\gamma_p\rightarrow1$,
one finds affine deformations with
$u_{p\alpha}(\lambda)=\lambda_{\alpha}u_{p\alpha}(\lambda=1)$.

Concerning the elastic properties, eq~\ref{eq:g} takes the form
\begin{equation}
  g_B(\lambda) =
  \frac{k_BT}{V}\sum_p
  \left( \frac{l_B(\lambda)}{k_p + l_B(\lambda)}\right)^2 .
\end{equation}
while the the shear modulus can be written as
\begin{eqnarray}
  \label{eq:GB}
  G_{B} &=& \frac{k_BT}{V}\sum_p \gamma_p^2.
\end{eqnarray}
Note the different functional form of eqs~\ref{eq:GA} and~\ref{eq:GB}.
Since $0\le \gamma \le 1$, the contribution of confined modes to the
elastic response is stronger in Model A than in Model B. Furthermore,
within Model B the interplay between the network connectivity
(represented by the eigenmode spectrum $\{k_p\}$ of the Kirchhoff
matrix) and the confining potential $l$ is different for the shear
modulus eq~\ref{eq:GB} and the tube diameter eq~\ref{eq:dT}.

\subsection{Model C: Simultaneous presence of both types of confinement}

Finally, we can discuss a situation where confinement effects of type
A and B are present simultaneously. Coupling each node to
two extra springs $l_A(\lambda)=l_A$ and 
$l_{B\alpha}(\lambda)=l_{B\alpha}/\lambda_\alpha^2$
leads to the following Hamiltonian in the eigenmode
representation:
\begin{equation}
  {\cal H}_p = \frac{k_p}{2}u_p^2 + \frac{l_A}{2}(u_p-v_{Ap}(\lambda))^2 +
  \frac{l_B(\lambda)}{2}(u_p-v_{Bp}(\lambda))^2.
\end{equation}
Model A and Model B are recovered by setting  $l_A$ respectively $l_B$
equal to zero. Furthermore we assume, that both types of confinement can
be activated and deactivated independently. This requires,

\begin{eqnarray}
  \langle v_{Ap}^2\rangle &=& \frac{l_A+k_p}{l_A k_p}
\label{eq:vpAC}\\
  \langle v_{Bp}^2\rangle &=& \frac{l_B+k_p}{l_B k_p}.
\label{eq:vpBC}
\end{eqnarray}
In the presence of both types of confinement, the mean excitation
of the modes is given by
\begin{eqnarray}
  \langle U_p^2\rangle(\lambda) &=& 
\frac{l_A^2\ \langle v_{Ap}^2\rangle(\lambda)
    + l_B^2(\lambda)\ \langle  v_{Bp}^2\rangle(\lambda) } 
      {(l_A+l_B(\lambda)+k_p)^2} \nonumber\\
&+& \frac{2 l_A l_B(\lambda)\ \langle v_{Ap} v_{Bp}\rangle(\lambda)} 
         {(l_A+l_B(\lambda)+k_p)^2} \label{eq:UpC}
\end{eqnarray}
while the thermal fluctuations are reduced to
\begin{equation}
  \langle \delta u_p^2\rangle(\lambda) = \frac{k_BT}{l_A+l_B(\lambda)+k_p}.
\label{eq:deltaupC}
\end{equation}
Finally the condition that the simultaneous presence of both constraints does
not affect ensemble averages in the state of preparation requires
\begin{eqnarray}
  \langle v_{Ap}v_{Bp}\rangle &=& \frac{k_BT}{k_p}\ .
\label{eq:vpAvpBC}
\end{eqnarray}
From eqs~\ref{eq:vpAC} to~\ref{eq:vpAvpBC} one can
calculate  the deformation dependent total excitation of the modes:
\begin{equation}
  \begin{split}
    \frac{\langle u_p^2\rangle(\lambda)}{\langle u_p^2\rangle} &= 1 +
    (\lambda^2-1)\left(\frac{l_A}{k_p+l_A+\frac{l_B}{\lambda^2}}\right.\\
    &+
    \left. \frac{\frac{l_B}
        {\lambda^2}\left(l_A+\frac{l_B}{\lambda^2}\right)}
      {\left(k_p+l_A+\frac{l_B}{\lambda^2}\right)^2}\right),
  \end{split} \label{dist_dt}
\end{equation}
so that
\begin{equation}
  \label{eq:gC}
  g_C(\lambda) =
  \frac{k_BT}{V}\sum_p
  \left(\frac{l_A}{k_p+l_A+\frac{l_B}{\lambda^2}} +
    \frac{\frac{l_B}
      {\lambda^2}\left(l_A+\frac{l_B}{\lambda^2}\right)}
    {\left(k_p+l_A+\frac{l_B}{\lambda^2}\right)^2}\right) \ .
\end{equation}
In the present case, the shear modulus can be written as
\begin{eqnarray} 
  \label{eq:GC}
  G_{C} &=& \frac{k_BT}{V}\sum_p 
  \frac{\gamma_{Ap} (1-\gamma_{Bp})}{1-\gamma_{Ap}\gamma_{Bp}}
  \nonumber\\
  && +
  \frac{\gamma_{Bp} (1-\gamma_{Ap}) \left( \gamma_{Bp} (1-\gamma_{Ap}) +
      \gamma_{Ap} (1-\gamma_{Bp})
    \right)}
  {\left(1-\gamma_{Ap}\gamma_{Bp}\right)^2} \ .\nonumber\\
\end{eqnarray}
Note that the shear modulus is not simply the sum of the contributions
from the A and B confinements.  While eqs~\ref{eq:GA} and~\ref{eq:GB}
are reproduced in the limits $\gamma_{Bp}=0$ and $\gamma_{Ap}=0$
respectively, eq~\ref{eq:GC} reflects the fact that a mode can never
contribute more than $k_BT$ to the shear modulus. Thus, for
$\gamma_{Bp}=1$ (respectively $\gamma_{Ap}=1$) the $p$th mode
contributes this maximum amount independently of the value of
$\gamma_{Ap}$ (respectively $\gamma_{Bp}$).

An important point, which holds for all three models, is that it is
{\em not} possible to estimate the confinement contribution to the
shear modulus from the knowledge of the absolute strength $l_A,l_B$ of
the confining potentials alone. Required is rather the knowledge of
the {\em relative} strengths $\gamma_{Ap},\gamma_{Bp}$ which in turn
are functions of the network connectivity.

\subsection{Discussion}
It is not a priori clear, whether entanglement effects are more
appropriately described by model A or model B. While model A has the
benefit of simplicity, Ronca and Allegra proposed model
B,~\cite{RoncaAll_jcp_75} because it leads (on length scales beyond
the tube diameter) to the conservation of intermolecular contacts
under strain.  Similar conclusions were drawn by Heinrich and
Straube~\cite{Heinrich_advpol_88} and Rubinstein and
Panyukov.~\cite{RubinPan_mm_97} In the end, this problem will have to
be resolved by a {\em derivation} of the tube model from more
fundamental topological considerations. For the time being, an
empirical approach seems to be the safest option.  Fortunately, the
evidence provided by experiments~\cite{Straube_prl_98} and by
simulations~\cite{RE_njp_99} points into the same direction.

\begin{figure}  
  \begin{center}
  \includegraphics[angle=0,width=8.25cm]{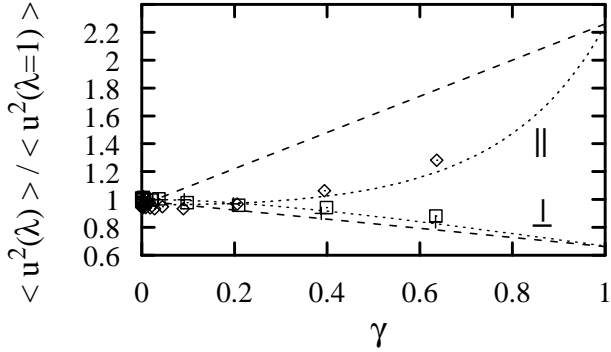}
  \end{center} 
  \caption{Excitation of constrained modes parallel and
    perpendicular to the elongation at $\lambda=1.5$ as a function of
    the mode degree of confinement $0\le\gamma\le1$. The dashed (dotted)
    lines show the predictions eq~\protect\ref{eq:u2_lambda_A} of
    Model A and eq~\protect\ref{eq:u2_lambda_B} of Model B
    respectively for generalized Rouse modes of a phantom network with
    identical connectivity.  The symbols represent the result of computer
    simulations of defect-free model polymer
    networks.~\protect\cite{RE_njp_99} The investigated modes are
    single-chain Rouse modes for network strands of length $N\approx
    1.25 N_e$.}
  \label{fig0}
\end{figure}

Since details of the interpretation of the relevant experiments are
still controversial (see Section~\ref{sec:lozenges}), we concentrate
on simulation results where the strain dependence of approximate
eigenmodes of the phantom model was measured
directly.~\cite{RE_njp_99} Figure~\ref{fig0} shows a comparison of
data obtained for defect-free model polymer networks to the
predictions eq~\ref{eq:u2_lambda_A} of Model A and
eq~\ref{eq:u2_lambda_B} of Model B. The result is unanimous.  We
therefore believe eq~\ref{eq:l_AffTrans} and Model B to be the
appropriate choice for modeling confinement due to entanglements.  The
shear modulus of an entangled network should thus be given
by:~\cite{RE_epjb_98}

\begin{equation}
G = G_{ph} + \frac{k_BT}{V}\sum_p \gamma_p^2
\end{equation}
where in contrast to Reference~\cite{RE_epjb_98} the various $\gamma_p$
are no longer free parameters but depend through
eq~\ref{eq:gamma_p} on a single parameter: the strength $l$ of the
confining potential, which is {\em assumed} to be homogeneous for all
monomers.  The difficulty of this formal solution of the generalized
constrained fluctuation model for polymer networks is hidden in the
use of the generalized Rouse modes of the phantom model, which are
difficult to obtain for realistic
connectivities.~\cite{KlocMark_mm_90,Sommer_jcp_93} A useful ansatz
for end-linked networks is a separation into independent
Flory-Einstein respectively Rouse modes for the crosslinks and network
strands.~\cite{RE_njp_99,RE_epjb_98} In fact, the simulation results
presented in Figure~\ref{fig0} are based on such a decomposition.

For randomly crosslinked networks with a typically exponential
strand length polydispersity, the separation into Flory-Einstein 
and single-chain Rouse modes ceases to be useful. In this case,
we can think of two radically different strategies:
\begin{itemize}
\item To keep the network connectivity in the analysis. For example,
  there is no principle reason, why the methods presented by Sommer et
  al.~\cite{Sommer_jcp_93} and Everaers~\cite{RE_njp_99} could not be
  combined, in order to investigate the strain dependence of
  constrained generalized Rouse modes in computer simulations. Note,
  however, that this completely destroys the self-averaging properties
  of the approximation used in Reference~\cite{RE_njp_99}. Analytic
  progress in the evaluation of, for example, eq~\ref{eq:GB} for the
  entanglement contribution to the shear modulus requires information
  on the statistical properties of the eigenvalue spectra of networks
  generated by random crosslinking.  To our knowledge, the only
  available results were obtained numerically by Shy and
  Eichinger.~\cite{ShyEichinger_jcp_89} Note, that model C is
  irrelevant, {\em if one is able to carry out calculations with the
    proper network eigenmodes}.
\item To average out the connectivity effects in tube models for
  polymer networks.~\cite{WarnerEdwards_jpa_78} In the second part of
  the paper, we will consider {\em linear} chains under the influence
  of two types of confinement: network connectivity and entanglements.
\end{itemize}

\section{Tube Models}

In SANS experiments of dense polymer melts, it is possible to measure
single chain properties by deuterating part of the
polymers.\cite{benoit} If such a system is first crosslinked into a
network and subsequently subjected to a macroscopic strain, one can
obtain information on the microscopic deformations of labeled random
paths through the network.\cite{benoit} In order to interpret the
results, they need to be compared to the predictions of theories of
rubber elasticity. Unfortunately, for randomly crosslinked networks it
is quite difficult to calculate the relevant structure factors even in
the simplest
cases.~\cite{HiggsBall_jpf_88,Cloizeaux_jp_94,Ullman_mm_79} Because
the crosslink positions on different precursor chains should be
uncorrelated, Warner and Edwards~\cite{WarnerEdwards_jpa_78} had the
idea to consider a tube model, where the crosslinking effect is
``smeared out'' along the chain. To model confinement due to
crosslinking, they used (in our notation) Model A, since this ansatz
reproduces the essential properties of phantom models (affine
deformation of equilibrium positions and deformation independence of
fluctuations). In contrast, Heinrich and
Straube~\cite{Heinrich_advpol_88}, and Rubinstein and
Panyukov~\cite{RubinPan_mm_97} treated confinement due to
entanglements using Model B. Obviously, both effects are present
simultaneously in polymer networks.  In the following, we will develop
the idea that in order to preserve the qualitatively different
deformation dependence of the two types of confinement, they should be
treated in a ``double tube'' model based on our Model C.

Before entering into a detailed discussion, we would like to point out
a possible source of confusion related to the ambiguous use of the
term ``tube'' in the literature (including the present paper).  A real
tube is a hollow, cylindrical object, suggesting that in the present
context the term should be reserved for the {\em confining potential}
described by quantities such as $\vec \xi_i, \vec v_p, l$. It is in
this sense that we speak of an ``affinely deforming tube''. However, a
harmonic confining tube potential is a theoretical construction which
is difficult to visualize. For example, in the continuum chain limit
used below, the forces exerted ``per monomer'' become infinitely small
corresponding to $\vec \xi_i \rightarrow \infty, l\rightarrow 0$.  On
the other hand, the term tube is often associated with the tube
``contents'', i.e. the superposition of the accessible polymer
configurations characterized via a locally smooth tube axis (the
equilibrium positions $\vec U_p$) and a tube diameter $d_T$ (defined
via the fluctuations $\vec{\delta u}_p$). This second definition
refers to measurable quantities.\cite{benoit} Which kind of tube we
are referring to, will hopefully always be clear from the context and
the mathematical definition of the objects under discussion.

In the case of linear polymers, the phantom model reduces to the Rouse
model with vanishing equilibrium positions $\vec R_i \equiv 0$. As a
consequence, there are no strain effects other than those caused by
the confinement of thermal fluctuations. In particular, the ``intrinsic''
phantom modulus vanishes (see eq~\ref{eq:Fph}). Since
the networks are modeled as superpositions of independent linear
paths, we have to introduce confinement of type A in order to recover
the phantom network shear modulus $G_{ph}$ in the absence of entanglements.

In the Rouse model the Kirchhoff matrix takes the simple tridiagonal
form

\begin{equation}
  \Mtensor{K} = k\begin{pmatrix} -2 & 1 & 0
    &\cdots & & & 0 \\ 1 & -2 & 1 & 0 & \cdots & \\ & & \ddots & & & &
    \\  & & & & & & \\ & & & & & & \\ 0 &\cdots & & & & 1 & -2 \end{pmatrix}
\end{equation}
and, depending on the boundary conditions, is diagonalized by
transforming to sin or cos modes using the transformation matrix

\begin{equation}
  \Mtensor{S} = (S)_{jp} =
  \frac{1}{\sqrt{N}}\exp\left(\mbox{i}\pi\frac{jp}{N}\right).
\end{equation}
The eigenvalues of the diagonalized Kirchhoff matrix 
$(\MTtensor{K})_{pp} = (\Mtensor{S}^{-1}\Mtensor{K}\Mtensor{S})_{pp} = k_p$
are given by
\begin{equation}
  k_p = 4k\sin^2\left(\frac{p\pi}{2N}\right).
\end{equation}
If we consider a path with given radius of gyration $R_g^2$,
the basic spring constant is given by $k= \frac{N\,k_BT}{2R_g^2}$.
In the continuous chain limit ($N\rightarrow\infty$), 
sums over eigenmodes can be approximated
by integrals. For example,  
one obtains from eq~\ref{eq:dT} an expression
for the tube diameter
\begin{eqnarray}
  d_{T\alpha}^2 &=& \frac{1}{N} \int_0^N dp \,\frac{1}{k_p + l} \nonumber\\
  &=& \frac{k_BT}{{\sqrt{l\left( 4k + l \right) }}}\approx \frac{k_BT}{2\sqrt{lk}}
\end{eqnarray}
which could be further simplified, since in this limit
the springs representing a chain
segment between two nodes are much stronger than the springs realizing
the tube, i.e. $k\gg l$.

For normally distributed internal distances $\vec r_{xx'}$ between
points $x=\frac{n}{N}$, $x'=\frac{m}{N}$ on the chain contour the
structure factor is given by
\begin{equation} \label{eq:Sq}
  S(\vec{q},\lambda)
  = \int_{x=0}^1dx\int_{x'=0}^1dx'
  \exp\left(-\frac{1}{2}\sum_{\alpha=1}^3q^2_{\alpha}
    \langle r_{xx'\alpha}^2(\lambda)\rangle\right).
\end{equation}
In the present case, eq~\ref{eq:dist} reduces to
\begin{equation} \label{path_dist}
  \langle r^2_{xx'\alpha}(\lambda) \rangle = 
  \frac1N \int_{-\infty}^{\infty} dp \,\langle u^2_{p\alpha}(\lambda) \rangle 
  \left|e^{\mbox{i}\pi p x} - e^{\mbox{i}\pi p x'}\right|^2 \ .
\end{equation}
In the undeformed state,
\begin{equation}
  \langle r_{xx'\alpha}^2(\lambda=1)\rangle = 2 R_g^2 |x-x'|  
\end{equation}
so that the structure factor is given by the Debye function:
\begin{equation} 
  S(\vec{q},\lambda=1)
  = \frac{2N}{q^4R_g^4}\left(\exp(-q^2R_g^2)-1+q^2R_g^2\right) \ .
\end{equation}

\subsection{The Warner--Edwards Model} \label{sec:WarnerEdwards}
Warner and Edwards~\cite{WarnerEdwards_jpa_78} used the replica method
to calculate the conformational statistics of long paths through
randomly crosslinked phantom networks. The basic idea was to represent
the localization of the paths in space due to their integration into a
network by a coarse-grained tube-like potential.  Recently, it was
shown by Read and McLeish~\cite{ReadMcLeish_prl_97,ReadMcLeish_mm_97}
that the same result could be obtained along the lines of the
following, much simpler calculation, where we evaluate Model A for
linear polymers.

Evaluation of the integrals eqs~\ref{eq:dT}
and~\ref{eq:dist} yields for the deformation independent
tube diameter and the internal distances:

\begin{eqnarray} 
  \label{eq:dtA}
  d_{TA\alpha}^2 &=& \frac{k_BT}{2\sqrt{k l_A}}\\
  \label{eq:rxxA}
  \frac{ \langle r_{xx'\alpha}^2(\lambda)\rangle}{2 R_g^2} &=& 
  \lambda_{\alpha}^2|x-x'|\\
  &+&\left(1-\lambda_{\alpha}^2\right)\frac{d_{TA\alpha}^2}{R_g^2}
  \left(1-e^{-\frac{R_g^2|x-x'|}{d_{TA\alpha}^2}}\right),
  \nonumber
\end{eqnarray}
We note that the latter equation can be rewritten in the form

\begin{equation}
\begin{split}
  \label{eq:f_A}
  \frac{\langle r_{xx'\alpha}^2 (\lambda) \rangle - \langle
    r_{xx'\alpha}^2 (1) \rangle}{(\lambda_\alpha^2-1) \langle
    r_{xx'\alpha}^2 (1) \rangle} 
  &= f_A\left(\frac{R_g^2|x-x'|}{d_{TA}^2}  \right)\\
  f_A(y) &= 1+\frac{e^{-y}-1}y
\end{split}
\end{equation}
with a universal scaling function $f_A(y)$ which does {\em not} depend
explicitly on the deformation. 
eq~\ref{eq:f_A} measures the degree of affineness of 
deformations on different length scales. Locally, i.e.
for distances inside the tube with $R_g^2|x-x'| \ll d_{T}^2$
corresponding to $y\ll 1$, the polymer remains undeformed. Thus
$\lim_{y\rightarrow 0} f(y) = 0$. Deformations become
affine for $R_g^2|x-x'| \gg d_{T}^2$
and $y\gg 1$, where $f(y)$ tends to one.

Furthermore, one obtains for the shear modulus 
and the stress-strain relation
\begin{eqnarray} 
  g_A(\lambda) &=& \frac{\rho b^2 \sqrt{k l_A}}{6} = G_A\\ 
  \label{eq:G_A}
  G_{A}  &=& \frac{1}{4}\frac{\rho b^2k_BT}{d_{TA}^2}\\ 
  \sigma_T(\lambda) &=& (\lambda^2-\frac{1}{\lambda})G_A.
\end{eqnarray}
so that the Mooney-Rivlin parameters are simply given by
\begin{eqnarray} 
  2 C_1 &=& G_{A} \\
  2 C_2 &=& 0 \ .
\end{eqnarray}

\subsection{The Heinrich--Straube / Rubinstein--Panyukov-Model}
Heinrich and Straube,~\cite{Heinrich_advpol_88} and Rubinstein and
Panyukov~\cite{RubinPan_mm_97} have carried out analogous
considerations for Model B, i.e. an affinely deforming tube. The
relation between the strength of the springs $l_B$ and the tube
diameter in the unstrained state is identical to the previous case.
However, the tube diameter now becomes deformation dependent:

\begin{eqnarray}
 \label{eq:dTB}
  d_{TB\alpha}^2(\lambda) &=& \lambda_{\alpha} \frac{d_{TB}^2}3 \ .
\end{eqnarray}
Thus the typical width of the fluctuations changes only
with the square root of the width of the confining potential.
Using equations eqs~\ref{path_dist} and~\ref{eq:u2_lambda_B}, 
one obtains for the mean square internal distances:

\begin{equation} 
\label{eq:rxxB}
  \begin{split}
    \frac{ \langle r_{xx'\alpha}^2(\lambda)\rangle}{2 R_g^2} &=
    \lambda_{\alpha}^2|x-x'|\\
    &+\frac{1}{2}\left(\lambda_{\alpha}^2-1\right)|x-x'|
      e^{-\frac{R_g^2|x-x'|}{d^2_{TB\alpha}(\lambda)}}\\
    &-\left. \frac
      32\left(\lambda_{\alpha}^2-1\right)\frac{d^2_{TB\alpha}(\lambda)}{R_g^2}
      \left(1-
        e^{-\frac{R_g^2|x-x'|}{d^2_{TB\alpha}(\lambda)}}\right)\right. \ .
  \end{split}
\end{equation}
Again, we can rewrite this result in terms of a universal
scaling function for the degree of affineness of the polymer
deformation:

\begin{equation}
\begin{split}
  \label{eq:f_B}
  \frac{\langle r_{xx'\alpha}^2 (\lambda) \rangle - \langle
    r_{xx'\alpha}^2 (1) \rangle}{(\lambda_\alpha^2-1) \langle
    r_{xx'\alpha}^2 (1) \rangle} 
  &= f_B\left(\frac{R_g^2|x-x'|}{d^2_{TB\alpha}(\lambda)}
  \right) \\
  f_B(y) &= 1+\frac12e^{-y}+ \frac32 \frac{e^{-y}-1}y \ .
\end{split}
\end{equation}
Eq~\ref{eq:f_B} shows that Straube's
conjecture~\cite{Straube_mm_94,Straube_prl_95,Straube_mm_96}
$f_A(y)=f_B(y)$ is incorrect. However, the two functions are
qualitatively very similar.

For the shear modulus and the stress-strain relation we find

\begin{eqnarray}
 \label{eq:gB}
 g_{B}(\lambda) &=& \frac{1}{8}\frac{\rho b^2k_BT}{\lambda d_{TB}^2}
 = \frac{G_B}{\lambda}\\
 \label{eq:G_aff}
 G_{B} &=& \frac{1}{8}\frac{\rho b^2k_BT}{d_{TB}^2}\\
 \label{eq:sigma_aff}
 \sigma_T(\lambda) &=&
 \left(\sqrt{\lambda} - \frac{1}{\sqrt{\lambda}} +
   \lambda-\frac{1}{\lambda}\right)G_{B}
\end{eqnarray}
in agreement with Rubinstein and Panyukov.~\cite{RubinPan_mm_97} In
order to account for the network contribution to the shear modulus,
these authors add the phantom network results to
eqs~\ref{eq:G_aff},~\ref{eq:sigma_aff}.  This leads to the following
relations for the Mooney-Rivlin parameters:~\cite{RubinPan_mm_97}

\begin{eqnarray} 
  2 C_1 &=& G_{ph}+\frac12 G_B \\
  2 C_2 &=& \frac12 G_B \ .
\end{eqnarray}

Note that eq~\ref{eq:sigma_aff} holds only for $\lambda\approx1$.
For large compression or extension the approximation $k\gg l(\lambda)$
breaks down and one regains the result of Heinrich and
Straube:~\cite{Heinrich_advpol_88}

\begin{eqnarray}
\sigma_T(\lambda) &=&
 \left(\lambda - \frac{1}{\sqrt{\lambda}}\right)G_{B}
\ \ \ (\lambda\ll1,\ \lambda\gg1) \ .
\end{eqnarray}

\subsection{The ``double tube'' model}

In the following we discuss a combination of two different
constraints, one representing the network (Model A) and therefore deformation
independent and the other representing the entanglements (Model B).
Thus we use Model C to combine the Warner-Edwards model with
the Heinrich-Straube/Rubinstein-Panyukov model. 

Evaluating eq~\ref{eq:dT} one obtains for the tube diameter

\begin{eqnarray}
  \frac{1}{d_{TC\alpha}^4(\lambda)} &=& \frac{1}{d_{TA\alpha}^4} +
  \frac{1}{d_{TB\alpha}^{4}(\lambda)}\ \ .
\end{eqnarray}

The deformation dependent internal distances are
given by:
\begin{equation}
\label{eq:rxxC}
  \begin{split}
    \frac{ \langle r_{xx'\alpha}^2(\lambda)\rangle}{2 R_g^2} &=
    \lambda_{\alpha}^2|x-x'|\\
    &+\frac12 \left( \lambda^2_\alpha-1 \right)|x-x'|
    \frac{d^4_{TC\alpha}(\lambda)}{d^4_{TB\alpha}(\lambda)}
    e^{-\frac{R_g^2|x-x'|}{d^2_{TC\alpha}(\lambda)}} \\
    &-\frac32 \left(\lambda^2_{\alpha} -1 \right) \frac{
      d^2_{TC\alpha}(\lambda)}{R_g^2} \left( 1 -
      e^{-\frac{R^2_g|x-x'|}{{{d^2_{{TC\alpha}}}}(\lambda)}} \right)\\
    & \quad \frac{ d^4_{TC\alpha}(\lambda)\ \left( d_{TA\alpha}^4 +
        \frac23\, d^4_{TB\alpha}(\lambda)\right)}
    {d^4_{TB\alpha}(\lambda)\ d^4_{TA\alpha}} \ .
  \end{split}
\end{equation}
In this case, it is not possible to rewrite the result 
in terms of a universal scaling function, because the relative
importance of the two types of confinement is deformation 
dependent. Introducing 
$\Phi(\lambda)=d_{TC\alpha}^4(\lambda)/d_{TB\alpha}^4(\lambda)$,
eq~\ref{eq:rxxC} can be rewritten as

\begin{eqnarray}
  \label{eq:f_C}
  \lefteqn{
    \frac{\langle r_{xx'\alpha}^2 (\lambda) \rangle - \langle
      r_{xx'\alpha}^2 (1) \rangle}{(\lambda_\alpha^2-1) \langle
      r_{xx'\alpha}^2 (1) \rangle}
    \left(\frac{R_g^2|x-x'|}{d_{TC\alpha}^2(\lambda)}, \Phi(\lambda)
    \right) =}
  \hfill\nonumber\\ 
  & &  f_A\left(y\right)
  + \Phi(\lambda) \left(f_B\left(y\right)
    -f_A\left(y  \right)
  \right) \ .
\end{eqnarray}

For the elastic properties of the double tube model we find:

\begin{eqnarray}
  g_{C}(\lambda) &=& 
  \frac{2g_B^2(\lambda)+g_A^2}{\sqrt{4g_B^2(\lambda)+g_A^2}}\\ 
  \label{eq:sigmaC}
  \sigma_T(\lambda) &=& 
  \left( \lambda^2 - 1\right) g_C(\lambda) +
  \left( 1-\lambda^{-1}\right) g_C(\lambda^{-1/2}) \ .
\end{eqnarray}
Again, eq~\ref{eq:sigmaC} only holds for moderate strains.
Shear modulus and the Mooney-Rivlin
parameters are given by:

\begin{eqnarray}
  \label{GC}
  G_{C} &=& \frac{2G_B^2+G_A^2}{\sqrt{4G_B^2+G_A^2}}\\
 2C_1   &=& \frac{G_A^4+6G_B^2 G_A^2+4G_B^4}{\left(4G_B^2+G_A^2\right)^{3/2}}\\
 2C_2   &=& \frac{4G_B^4}{\left(4G_B^2+G_A^2\right)^{3/2}} \ .
\end{eqnarray}

\subsection{Comparison of the different tube models}

In the following we compare the predictions of the different models
for the microscopic deformations and the macroscopic 
elastic properties from two different points of view:
\begin{enumerate}
\item As a function of the network connectivity, i.e. the 
ratio of the average strand length $N_c$ between crosslinks
to the melt entanglement length $N_e$.
For this purpose, we identify $G_A$ with the shear modulus
of the corresponding phantom network $G_{ph}$:

\begin{eqnarray}
\frac{1}{4}\frac{\rho b^2k_BT}{d_{TA}^2} = G_A &=& 
G_{ph} =
\left(1-\frac2f\right) \frac{\rho k_BT}{N_c}\\
d_{TA}^2 &=& \frac{f-2}{4f} b^2 N_c
\end{eqnarray}
where we use $f=4$ for our plots.
Similarly, we choose for $G_B$ a value of the order of the melt plateau
modulus $G_e$:

\begin{eqnarray}
\frac{1}{8}\frac{\rho b^2k_BT}{d_{TB}^2}= G_B &=& G_e = 
\frac{3}{4}\frac{\rho b^2k_BT}{N_e}\\
d_{TB}^2 &=& \frac16 b^2 N_e \ .
\end{eqnarray} 
\item Assuming that the system is characterized by a certain
tube diameter $d_{TC}$ or shear modulus $G_C$, we discuss its response
to a deformation as a function of the relative importance $0\le
\Phi\le1$ of the crosslink and the entanglement contribution to the
confinement:

\begin{eqnarray}
  \Phi &=& \frac{d_{TC}^4}{d_{TB}^{4}}\\
  1-\Phi &=& \frac{d_{TC}^4}{d_{TA}^{4}}
\end{eqnarray}
where $\Phi$ is of the order $\left(1+\left(\frac{N_e}{N_c}\right)^2\right)^{-1}$ \ .
\end{enumerate}

\subsubsection{Elastic properties}

\begin{figure}  
  \begin{center}
    \includegraphics[angle=0,width=8.25cm]{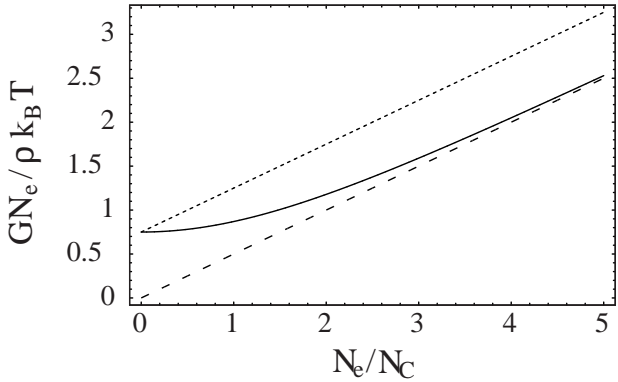}
  \end{center}
  \caption{Langley plot of the shear modulus. The solid line
    corresponds to the ``double tube'' model, the dotted line to
    the Heinrich--Straube/ Rubinstein--Panyukov model and the
    dashed line to the phantom model. $N_e$ represents the
    entanglement length and $N_c$ the crosslink length.}
  \label{fig1}
\end{figure}

\begin{figure}  
  \begin{center}
    \includegraphics[angle=0,width=8.25cm]{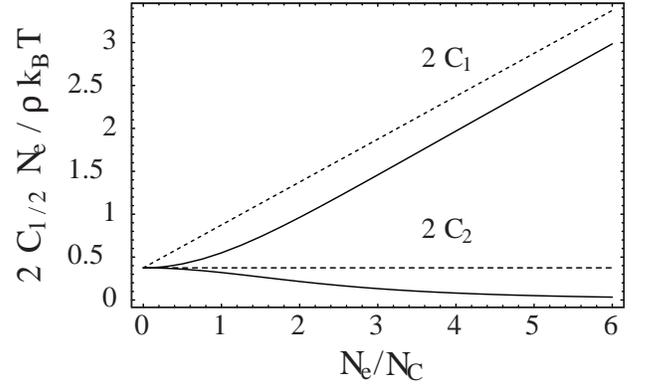}
  \end{center}
  \caption{Plot of the parameters $2 C_1$ and $2 C_2$ of the
    Mooney-Rivlin equation $f\left(\lambda^{-1}\right) = 2 C_1 + 2
    C_2\lambda^{-1}$ for the Rubinstein--Panyukov model (dotted) and
    the ``double tube'' model (solid).}
  \label{fig2}
\end{figure}

Figure~\ref{fig1} shows the shear modulus dependence on the
ratio of the network strand length $N_c$ to the melt entanglement
length $N_e$. 
As expected $G_C$ crosses over from $G_{ph}$ for short strands to
$G_e$ in the limit of infinite strand length. For comparison we have
also included the prediction of Rubinstein and Panyukov,
$G_{ph}+G_e$. The shear moduli predicted by our ansatz are always
smaller than this sum.  In particular, we find $G=G_{ph}$ for $N_c\ll
N_e$. The physical reason is that in a highly crosslinked network the
typical fluctuations are much smaller than the melt tube diameter. As
a consequence, the network does not feel the additional confinement
and the entanglements do not contribute to the elastic response.
Figure~\ref{fig2} shows analogous results for
the Mooney-Rivlin Parameters $C_1$ and $C_2$ again in comparison to
the predictions of Rubinstein and Panyukov. Note, that
$C_2$ is not predicted to be strand length independent.

\begin{figure}  
  \begin{center}
    \includegraphics[angle=0,width=8.25cm]{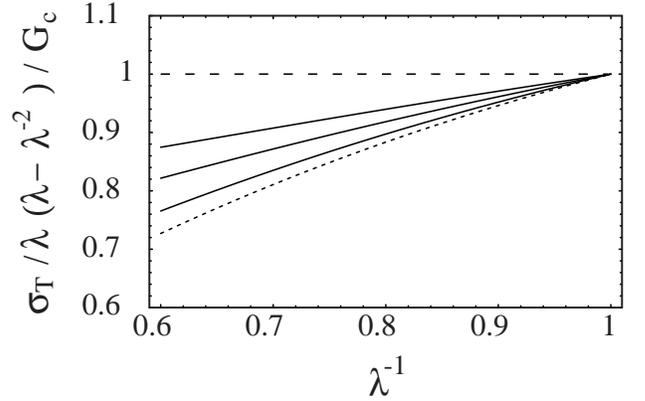}
  \end{center}
  \caption{Mooney-Rivlin representation of the reduced force for
    different values of $\Phi$ (from top to bottom: the Phantom model
    (dashed line, $\Phi=0$), the ``double tube'' model (solid lines,
    $\Phi=\frac13,\frac12,\frac34$) and the Heinrich-Straube/
    Rubinstein-Panyukov model (dotted line, $\Phi=1$)).}
  \label{fig3}
\end{figure}

Figure~\ref{fig3} shows the reduced force in the Mooney-Rivlin
representation for different entanglement contributions $\Phi$ to the
confinement.  For moderate elongations up to $\lambda\approx2$ the
curves are well represented by the Mooney-Rivlin form. For a given
shear modulus, $C_1$ and $C_2$ are a function of the entanglement
contribution $\Phi$ to the confinement:

 \begin{eqnarray}
 2C_1   &=& G_C \left(1-\frac{\Phi^2}{4-2\Phi}\right)\\
 2C_2   &=& G_C \frac{\Phi^2}{4-2\Phi} \ .
\end{eqnarray}

\subsubsection{The tube diameter}

Since eq~\ref{eq:GC} can be written in the form

 \begin{eqnarray}
  G_{C} &=& \frac{1}{8}\frac{\rho b^2k_BT}{d_{TC}^2}
   \left( 2-\Phi \right)\ \ ,
\end{eqnarray}
a plot of $d_{TC}^{-2}$ versus $N_e/N_c$ looks very similar to
Figure~\ref{fig1}.

\begin{figure}  
  \begin{center}
    \includegraphics[angle=0,width=8.25cm]{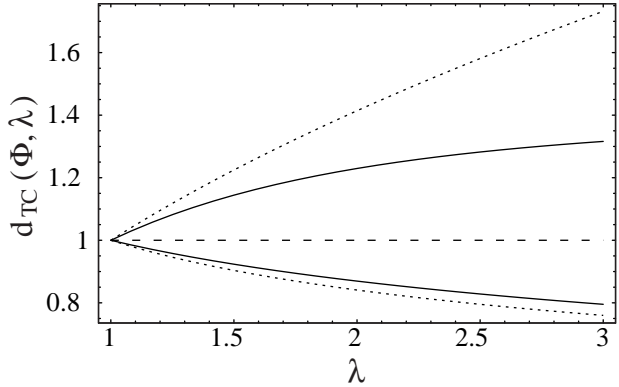}
  \end{center}
  \caption{Tube diameter $d_{TC}(\Phi,\lambda) =
    \left((1-\Phi)+\frac{\Phi}{\lambda^2}\right)^{-\frac{1}{4}}$ in
    parallel (upper curves) and perpendicular stretching direction for
    different elongation ratios $\lambda$ whereas $\Phi$ can be
    expressed by the entanglement length $N_e$ and the crosslink
    length $N_c$ by $\Phi = \frac{d_{TC}^4}{d_{TB}^4} =
    \frac{1}{1+\left(\frac{N_e}{N_c}\right)^2}$ using
    $\frac{d_{TB}^2}{d_{TA}^2} = \frac{N_e}{N_c}$. The dashed curve
    corresponds to the Warner--Edwards model, i.e.
    $d_{TC}(\Phi=0,\lambda)$, the dotted curve corresponds to the
    Heinrich--Straube/ Rubinstein--Panyukov model, i.e.
    $d_{TC}(\Phi=1,\lambda)$, and the solid line represents the
    ``double tube'' model with $\Phi=\frac3 4$.}
  \label{fig4}
\end{figure}

The deformation dependence of the tube diameter (Figure~\ref{fig4}) takes 
the form:

\begin{eqnarray}
  \frac{d_{TC\alpha}^4}{d_{TC\alpha}^4(\lambda)} &=& \left(1-\Phi\right) +
  \frac{\Phi}{\lambda_\alpha^2}\\
\lim_{\lambda\rightarrow\infty} \frac{d_{TC||}(\lambda)}{d_{TC||}} &=& 
\left(1-\Phi\right)^{-1/4}\\
\lim_{\lambda\rightarrow\infty} \frac{d_{TC\perp}(\lambda)}{d_{TC\perp}} &=& 
\left(\Phi\lambda\right)^{-1/4} \ .
\end{eqnarray}
In the parallel direction, the entanglement contribution to the
confinement vanishes for large $\lambda$ so that
$\lim_{\lambda\rightarrow\infty} d_{TC||}(\lambda) = d_{TA||}$.  On
the other hand, the entanglements become relatively stronger in the
perpendicular direction with $\lim_{\lambda\rightarrow\infty}
d_{TC\perp}(\lambda) = d_{TB\perp}(\lambda)$.

\subsubsection{Microscopic deformations and structure functions}
\label{sec:lozenges}
\begin{figure}  
  \begin{center}
    \includegraphics[angle=0,width=8.25cm]{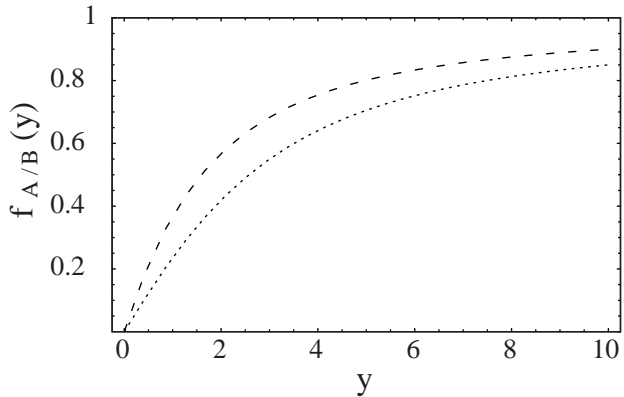}
  \end{center}
  \caption{Comparison of the universal scaling functions of
    eqs~\ref{eq:f_A},~\ref{eq:f_B} for the Warner-Edwards model
    (dashed) and the Heinrich--Straube/ Rubinstein--Panyukov model
    (dotted) with $y = \frac{R_g^2|x-x'|}{d_{TA/B}^2}$.}
  \label{fig5}
\end{figure}

Figure~\ref{fig5} compares the universal scaling functions of the
Warner--Edwards and Heinrich-Straube/Rubinstein-Panyukov model defined
by eqs~\ref{eq:f_A},~\ref{eq:f_B}.

More important for the actual microscopic deformations than the
difference between these two functions is the fact, that the distances
are scaled with the deformation dependent tube diameter. As a
consequence, deformations parallel to the elongation are {\em smaller}
in Model B than in Model A, while the situation is reversed in the
perpendicular direction. In the general case (eq~\ref{eq:f_C} of
Model C), the results are further complicated by the deformation
dependent mixing of the two confinement effects.  Nevertheless,
eqs~\ref{eq:f_A},~\ref{eq:f_B}, and~\ref{eq:f_C} should be
useful for the analysis of simulation data where real space distances
are directly accessible.

\begin{figure}  
  \begin{center}
    \includegraphics[angle=0,width=7.25cm]{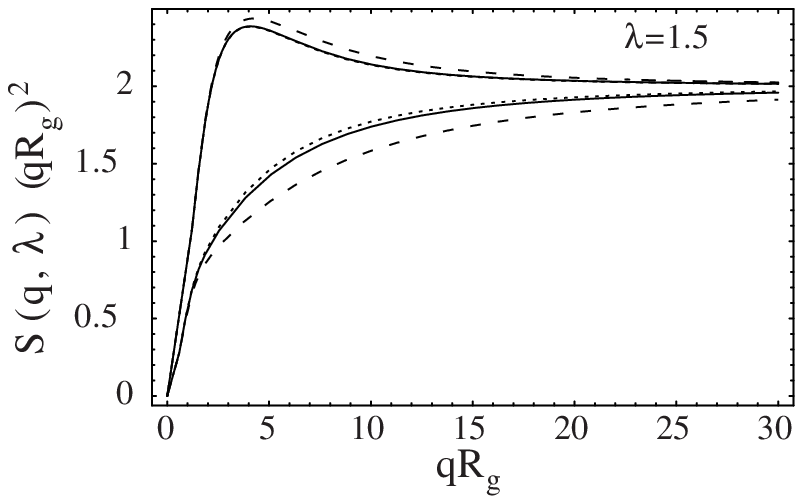}\newline
    \includegraphics[angle=0,width=7.25cm]{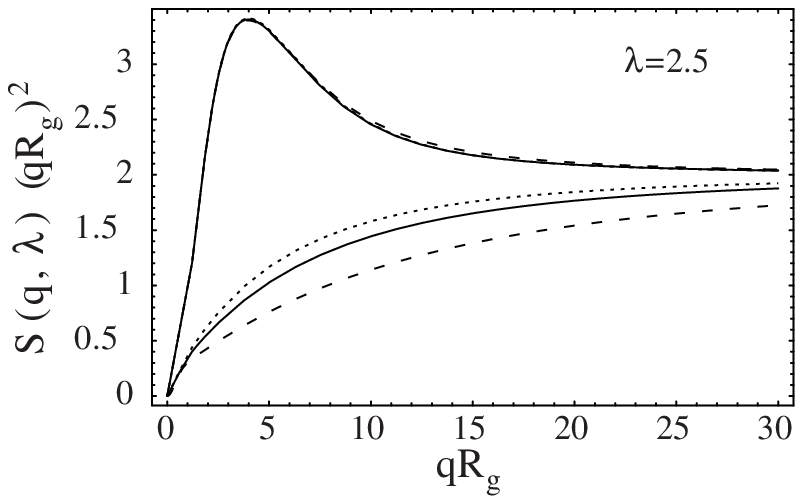}\newline
    \includegraphics[angle=0,width=7.25cm]{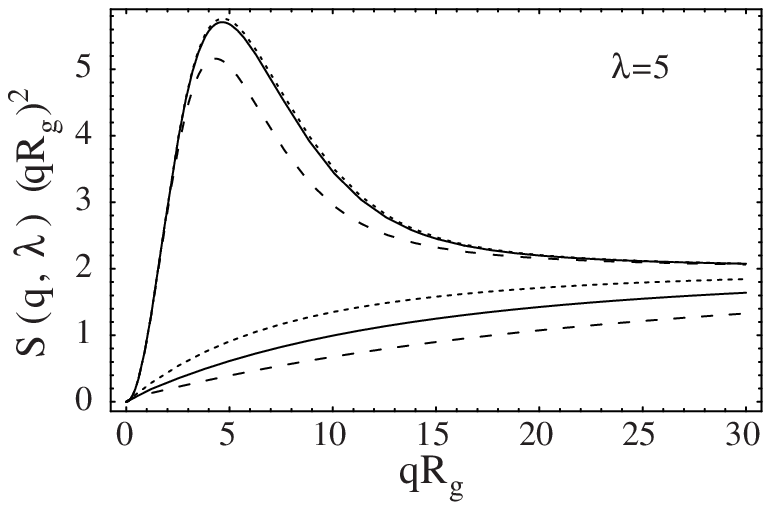}\newline
  \end{center}
  \caption{Kratky plots of the different structure factors in parallel
    and perpendicular stretching direction with $\frac{R_g}{d_T} = 6$:
    Warner--Edwards model (dashed), Heinrich--Straube/
    Rubinstein--Panyukov model (dotted line), ``double tube'' model
    with $\Phi = \frac 3 4$ solid line). The upper curves correspond
    to the perpendicular stretching direction.}
  \label{fig6}
\end{figure}

\begin{figure}  
  \begin{center}
    \includegraphics[angle=0,width=7.0cm]{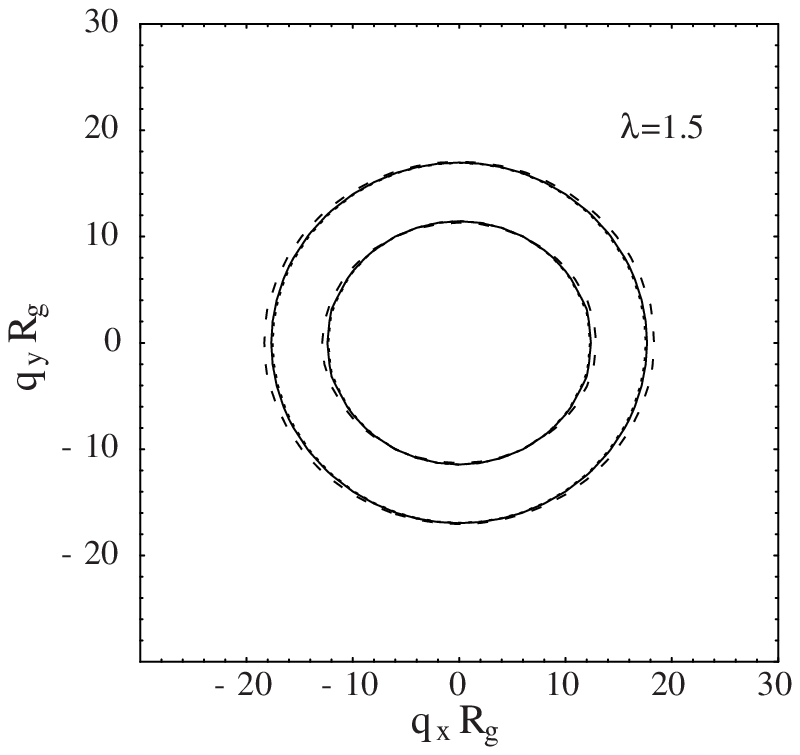}\\
    \includegraphics[angle=0,width=7.0cm]{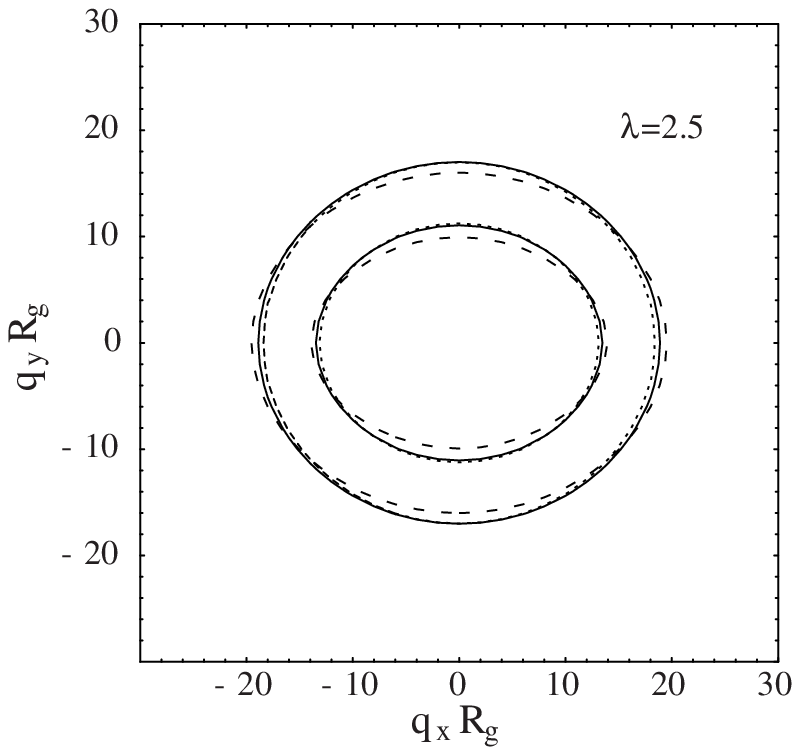}\\
    \includegraphics[angle=0,width=7.0cm]{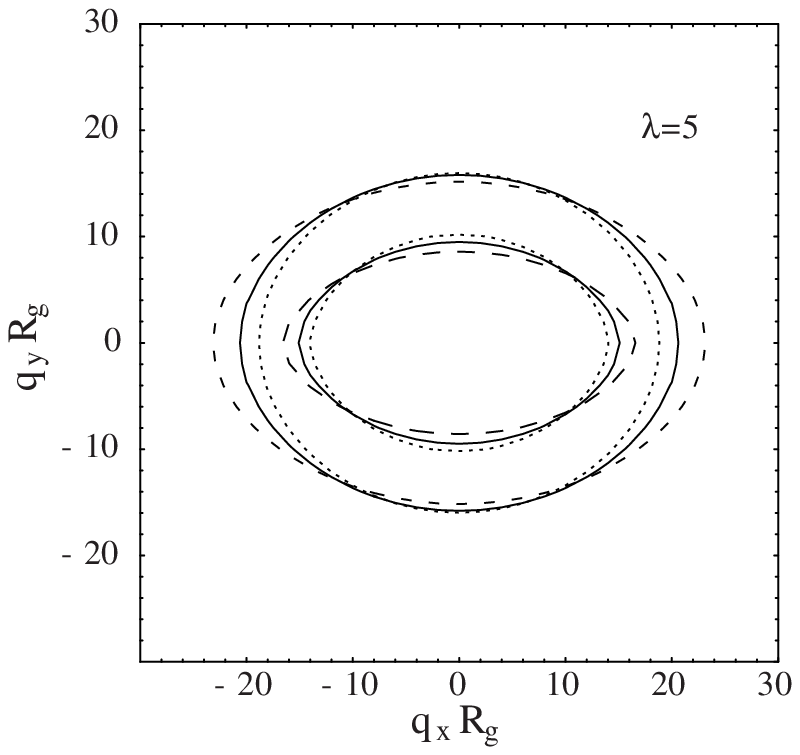}
  \end{center}
  \caption{Contour plots of the different structure factors with
    $\frac{R_g}{d_T} = 6$: Warner--Edwards model (dashed),
    Heinrich--Straube/ Rubinstein--Panyukov model (dotted line),
    ``double tube'' model with $\Phi = \frac 3 4$ (solid line). The
    upper curves correspond to the perpendicular stretching
    direction.}
  \label{fig7}
\end{figure}

Experimentally, the microscopic deformations can only be measured via
small angle neutron scattering.\cite{Straube_mm_94,Straube_prl_95}
Unfortunately, there seems to be no way to condense the structure
functions eq~\ref{eq:Sq} which result from
eqs~\ref{eq:rxxA},~\ref{eq:rxxB}, and~\ref{eq:rxxC} for different
strains into a single master plot. Figures~\ref{fig6} and~\ref{fig7}
show a comparison for three characteristic values of $\lambda$.
Qualitatively, the results for the three models are quite similar. In
particular, they do {\em not} predict Lozenge-like patterns for the
two-dimensional structure functions as they were observed by Straube
et al.~\cite{Straube_prl_95} In particular, we agree with Read and
McLeish~\cite{ReadMcLeish_prl_97} that the interpretation of Straube
et al.~\cite{Straube_mm_94,Straube_prl_95,Straube_prl_98} is based on
an ad-hoc approximation in the calculation of structure functions from
Model B. In principle, their alternative idea, to investigated the
influence of dangling ends on the structure function within Model A
and Model B,~\cite{ReadMcLeish_prl_97} can be easily extended to Model
C. Judging from the small differences between the models
(Figures~\ref{fig6} and~\ref{fig7}) and the results in
Ref.~\cite{ReadMcLeish_prl_97}, this would probably allow to obtain an
excellent fit of the data {\em and} to correctly account for the
deformation dependence of the tube.~\cite{Straube_prl_98} However,
since the lozenge patterns were also observed in tri-block systems
where only the central part of the chains was
labeled,~\cite{Straube_mm_96} dangling ends seem to be too simple an
explanation. At present it is therefore unclear, if the lozenge
patterns are a generic effect or if they are due to other artifacts
such as chain scission.~\cite{Straube_prl_98,Read_prl_98}
Simulations~\cite{RE_njp_99,Dueringkkgsg_Prl_91,DueringKKGSG_jcp_94,REKK_mm_95}
might help to clarify this point.

\section{Conclusion}
In this paper we have presented theoretical considerations related to
the entanglement problem in rubber-elastic polymer networks. More
specifically, we have dealt with constrained fluctuation models in
general and tube models in particular. The basic idea goes back to
Edwards,~\cite{Edwards_tube_procphyssoc_67} who argued that on a
mean-field level different parts of the network behave, as if they
were embedded in a deformation-dependent elastic matrix which exerts
restoring forces towards some rest positions. In the first part of our
paper, we were able to show that the generalized Rouse modes of the
corresponding phantom network without entanglement remain eigenmodes
in the presence of the elastic matrix.  In fact, the derivation of
eq~\ref{eq:HamConstMode}, which is the Hamiltonian of the exactly
solvable Constrained Mode Model (CMM),~\cite{RE_epjb_98} provides a
direct link between two diverging developments in the theory of
polymer networks: the ideas of Edwards, Flory and others on the
suppression of fluctuations due to entanglements and the
considerations of Eichinger,~\cite{Eichinger_mm_72}
Graessley,~\cite{Graessley_mm_75} Mark,~\cite{KlocMark_mm_90} and
others on the dynamics of (micro) phantom networks. An almost trivial
conclusion from our theory is the observation, that it is {\em not}
possible to estimate the entanglement effects from the knowledge of
the absolute strength of the confining potentials alone.  Required is
rather the knowledge of the {\em relative} strength which in turn is a
function of the network connectivity eq~\ref{eq:GB}.

Unfortunately, it is difficult to exploit our formally exact solution
of the constrained fluctuation model for arbitrary connectivity, since
it requires the eigenvalue spectrum of the Kirchhoff matrix for
randomly crosslinked networks. In the second part of the paper we have
therefore reexamined the idea of Heinrich and
Straube~\cite{Heinrich_advpol_88} to introduce entanglement effects
into the Warner-Edwards model~\cite{WarnerEdwards_jpa_78} for linear,
random paths through a polymer network, whose localization in space is
modeled by a harmonic tube-like potential. In agreement with Heinrich
and Straube,~\cite{Heinrich_advpol_88} and with Rubinstein and
Panyukov~\cite{RubinPan_mm_97} we have argued that in contrast to
confinement due to crosslinking, confinement due to entanglements is
deformation dependent. Our treatment of the tube model differs from
previous attempts in that we explicitly consider the simultaneous
presence of two different confining potentials. The effects are shown
to be non-additive.  From the solution of the generalized tube model
we have obtained expressions for the microscopic deformations and
macroscopic elastic properties which can be compared to experiments
and simulations.

While we believe to have made some progress, we do not claim to have
solved the entanglement problem itself. For example, it remains to be
shown how the geometrical tube constraint arises as a consequence of
the topological constraints on the polymer conformations. But even on
the level of the tube model, we are guilty of (at least) two possibly
important omissions: (i) we have neglected fluctuations in the local
strength of the confining potential and (ii) we have suppressed the
anisotropic character of the chain motion parallel and perpendicular
to the tube.  In the absence of more elaborate theories, computer
simulations along the lines of
Refs.~\cite{RE_njp_99,Dueringkkgsg_Prl_91,DueringKKGSG_jcp_94,REKK_mm_95,Sommer_jcp_93}
may present the best approach to a quantification of the importance of
these effects.

\subsection{Acknowlegements}
The authors wish to thank K. Kremer, M. P\"utz and T.A. Vilgis for helpful
discussions. We are particularly grateful to E. Straube for repeated
critical readings of our manuscript and for pointing out similarities
between our considerations and those by Read and McLeish.
\providecommand{\refin}[1]{\\ \textbf{Referenced in:} #1}

\newpage

\end{document}